\documentclass[useAMS,usenatbib,usegraphicx]{mn2e}
\usepackage[a4paper,centering, totalwidth=520pt, totalheight=700pt]{geometry}

\pdfoutput=1

\usepackage{aas_macros}
\usepackage{amssymb}
\usepackage{amsmath}
\usepackage{tabularx}
\usepackage{rotating}
\usepackage{bm}
\usepackage{multirow}
\usepackage{verbatim} 
\usepackage{color}

\begin{document}

\title[WiggleZ-BOSS overlap II: redshift-space distortions]{
The BOSS-WiggleZ overlap region II: dependence of cosmic growth on galaxy type}

\author[Mar\'in et al.]{\parbox[t]{\textwidth}{
Felipe A.\ Mar\'in$^{1,2}$\thanks{E-mail: fmarin@astro.swin.edu.au (FAM)}, Florian Beutler$^3$,
 Chris Blake$^{1,2}$, Jun Koda$^{1,2,4}$, Eyal Kazin$^{1,2}$, Donald P. Schneider$^{5,6}$}
\\ \\
$^1$ Centre for  Astrophysics \& Supercomputing, Swinburne University of Technology,
  P.O.\ Box 218, Hawthorn, VIC 3122, Australia\\
 $^2$ ARC Centre of Excellence for All-sky Astrophysics (CAASTRO), 44 Rosehill
St, Redfern, NSW 2016, Australia\\
$^3$ Lawrence Berkeley National Lab, 1 Cyclotron Rd, Berkeley, CA 94720, USA\\
$^4$ INAF - Osservatorio Astronomico di Brera, via Emilio Bianchi 46, I-23807 Merate, Italy\\
$^5$ Department of Astronomy \& Astrophysics, The Pennsylvania State University, University Park, PA 16802\\
$^6$ Institute for Gravitation and the Cosmos, The Pennsylvania State University, University Park, PA 16802
}

\date{\today, submitted to MNRAS}

\maketitle

\begin{abstract}
The anisotropic  galaxy 2-point correlation function (2PCF)  allows measurement of
the growth of large-scale structures from the effect of peculiar velocities
on the clustering pattern.
We present new
measurements of the auto- and cross- correlation function multipoles
of  69,180 WiggleZ and 
46,380 BOSS-CMASS galaxies sharing an overlapping volume of $\sim 0.2$ $(h^{-1}$Gpc)$^3$. 
Analysing the redshift-space distortions (RSD) of galaxy 2-point statistics for these two galaxy tracers, 
we  test for  systematic errors in the modelling depending on galaxy type 
 and investigate potential improvements in
cosmological constraints. 
We build a large number of mock galaxy catalogs to examine the limits of  different RSD models 
in terms of fitting scales and  galaxy type, and to study the covariance of the measurements when 
performing joint fits. 
For the galaxy data, fitting the monopole and quadrupole of the WiggleZ 2PCF on scales
 $24<s<80$ $h^{-1}$Mpc produces a measurement of the normalised growth rate  $f\sigma_8(z=0.54)=0.409\pm0.059$, whereas
for the CMASS galaxies we found a consistent constraint of $f\sigma_8(z=0.54)=0.466\pm0.074$, 
When combining the measurements,  
accounting for the correlation between the two surveys,
we obtain 
 $f\sigma_8(z=0.54)=0.413\pm0.054$, in  agreement with the $\Lambda$CDM-GR model of structure growth and 
 with other survey measurements.
  \end{abstract}


\section{Introduction}

The evolution of the spatial distribution of galaxies on large scales is deeply influenced by 
the physics of gravitational attraction, cosmic expansion, and the conditions of the early Universe, 
and constitutes an important probe and
discriminator of cosmological models. Spectroscopic galaxy surveys map this distribution using the 
distance-redshift relation, but due to peculiar velocities induced by the gravitational field,  
 these maps also contain 
`redshift-space distortions' (RSD) of the spatial positions of the galaxies, which 
modify the true (i.e. real space) pattern of the spatial clustering of galaxies.
\cite{kaiser:87} showed that on large scales 
the peculiar velocity field ${\bf v}$ (in dimensionless units of the Hubble velocity) is related to the matter overdensity $\delta_m$ as 
$\nabla\cdot{\bf v} =  -f\delta_m$, where the proportionality parameter $f(z)$ is called
the linear growth rate of structure. Modelling the redshift-space
clustering, in consequence,  allows us to constrain cosmological parameters through estimations
of $f(z)$.
Using Kaiser's findings, the pioneering 
 works in the 2dFGRS survey \citep{peacock_etal:01,hawkins_etal:03} in the local $z\sim0.1$ Universe,
measured the redshift-space two-point clustering of galaxies, which 
 resulted in a confirmation of the concordance $\Lambda$CDM model at present times. With the 
 advent of galaxy surveys at higher redshifts, we can now trace the history of $f(z)$ and obtain constraints 
 on cosmological models and the nature of dark energy \citep{linder_cahn:07}.
 
 However, important challenges must be addressed before we can use this tool 
effectively. 
On the observational side,
the most important factors limiting the statistical  precision 
of the clustering measurements  obtained from different surveys
 are the limited volume that the surveys can map, due to the 
 \emph{sample variance}
from fluctuations in the clustering on different regions of the universe, 
and the discreteness of the galaxy field known as 
\emph{shot noise} \citep[e.g.][]{kaiser:86,white_etal:09}

In addition, large-scale structures are 
 subject to a variety of systematic
 non-linear effects which affect our capacity to model the signal, 
 particularly on small scales.  First, we have 
non-linear growth of structure, such that  even on large scales,
the Kaiser relations are insufficient to account for the measured  clustering in galaxy data and
simulations.  
\cite{peacock:92}, and
 more recently \cite{scoccimarro:04, taruya_etal:10, seljak_mcdonald:11, wangl_etal:13}, 
 among others, have 
 improved the basic `Kaiser' model by including various non-linear effects in the matter clustering. 
 Second, there is  scale-dependent complexity in how galaxies trace haloes and cross-correlate to matter,
  known as {\it galaxy bias}. Third, galaxies possess non-linear pairwise velocities on small scales. 
The latest attempts to use the 2-point clustering pattern to model RSD have taken these and other effects into
account
\citep[e.g.][as recent examples]{reid_etal:12,beutler_etal:12rsd,sanchez_etal:13,delatorre_etal:13,contreras_etal:13,beutler_etal:14}, 
allowing us to confront predictions from different cosmological models.

\begin{figure*}
\includegraphics[width=0.48\linewidth]{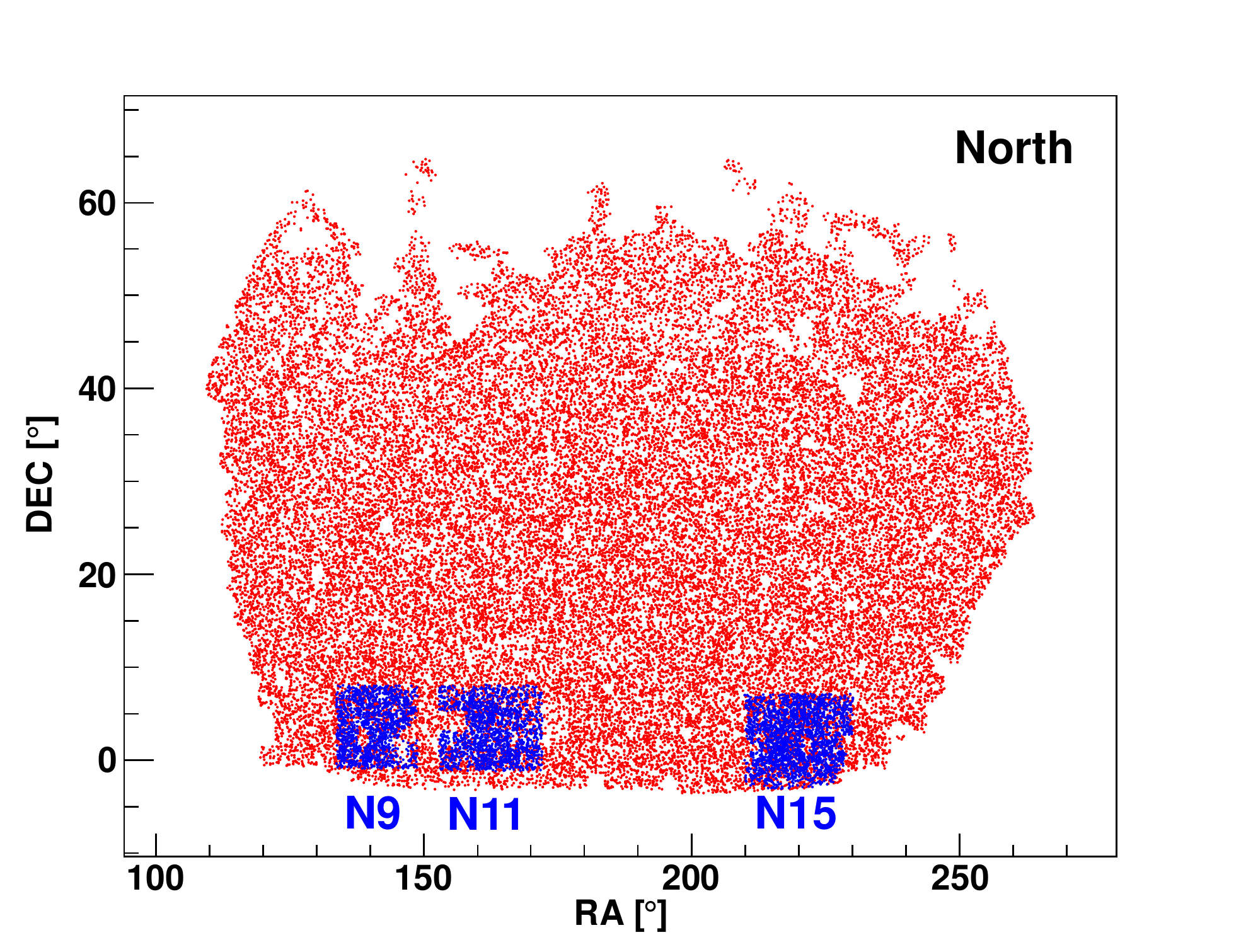}
\includegraphics[width=0.48\linewidth]{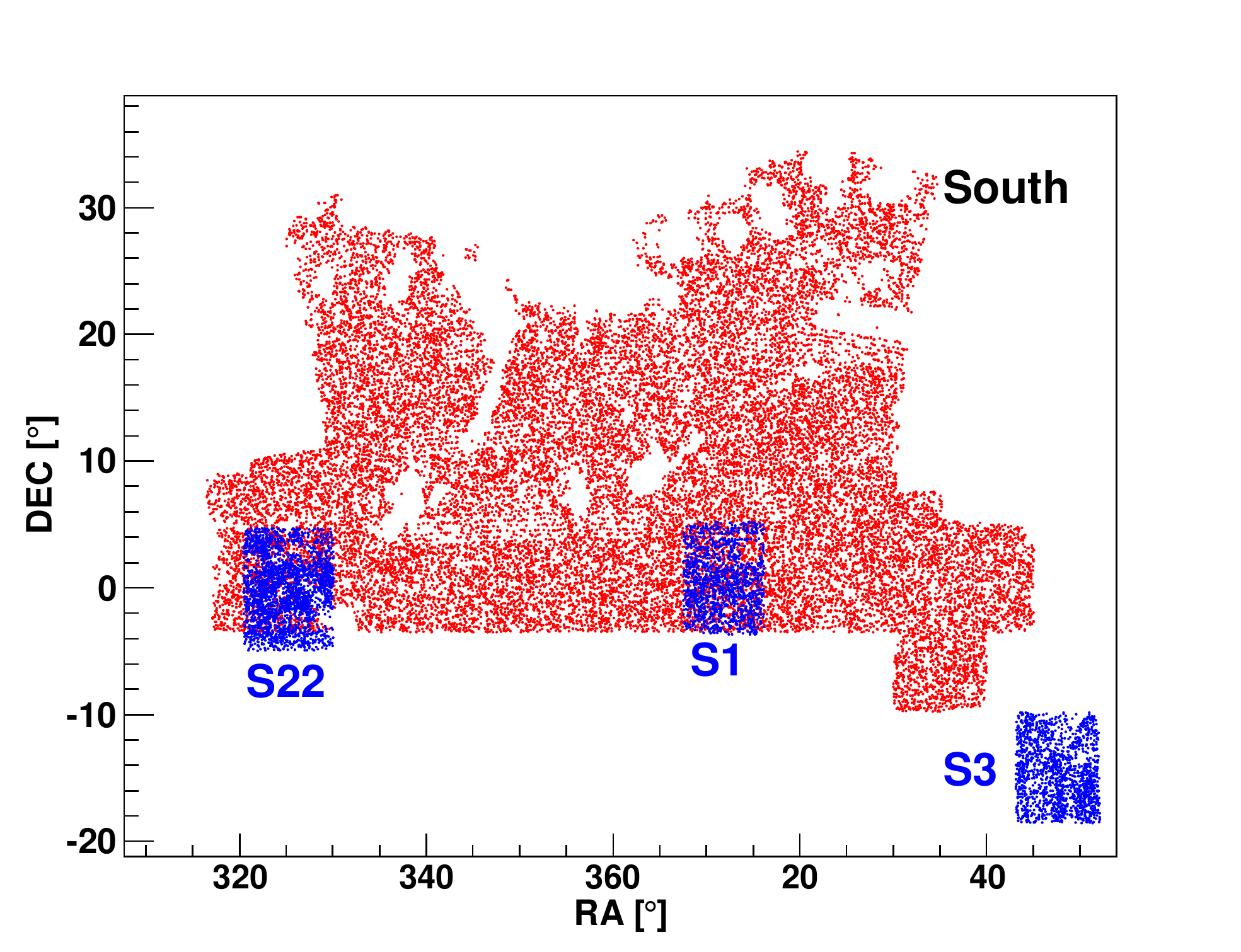}
\caption{\label{fig:allov}
 Sky coverage of BOSS-CMASS DR11 (red) and WiggleZ (blue). 
 The left panel shows the northern part of the surveys, while the right panel shows the southern sky coverage. 
 Five of the six WiggleZ regions are covered by CMASS, with a total of 69,180 WiggleZ galaxies and 46,380 
 CMASS galaxies in the overlap volume.}
\end{figure*}

Although in linear theory all galaxies respond as test particles to the
gravitational field, in detail the non-linear systematics depend on 
tracers themselves. This is because galaxy formation is affected by many non-linear
processes such as small-scale dynamics of halo formation, environment, and 
complex baryonic processes determining the
luminosity and colour at a given time, which are the main observables when selecting 
galaxies for a large-scale survey. 
Therefore, the analysis and modelling of two overlapping tracers makes it possible to constrain details of the clustering and
formation of the galaxy tracers themselves. Previous work has focused on  cross-correlating a tracer with well known 
properties with a second tracer we wish to study
\citep[e.g.][]{martinez_etal:99,chen:09, mountrichas_etal:09,font-ribera_etal:13}.
In our current study we approach this in  a cosmological context, 
in which a comparison of results using different tracers in the same
volume tests 
for systematic errors in  modelling of bias  and redshift-space distortions.

In this work we present measurements and  analysis of RSD using galaxies from the WiggleZ Dark Energy Survey 
\citep{drinkwater_etal:10} and the CMASS galaxy sample
from the Baryon Oscillation Spectroscopic Survey \cite[BOSS,][]{eisenstein_etal:11}. 
At a redshift of $z\approx 0.6$, the WiggleZ team targeted Emission Line 
Galaxies hosted in low-to-intermediate mass halos, which have low bias  
\citep[$b_{\rm WiggleZ}\sim 1$, see][]{blake_etal:11f, contreras_etal:13, marin_etal:13},
while the CMASS sample  consists of luminous, mostly red galaxies with  $b_{\rm CMASS}\sim 2$
\citep{reid_etal:12,tojeiro_etal:12,  chuang_etal:13, kazin_etal:13}
with similar number density $n\sim2-3\times10^{-4}$ ($h^{-1}$Mpc)$^{-3}$. 
With an overlap volume of approximately 0.2 ($h^{-1}$Mpc)$^{-3}$, this is, to date, the largest volume 
overlapping sample between two different galaxy redshift surveys.
We measure the redshift-space auto- and cross-
correlation functions of these galaxies and explore the constraints on the cosmic growth rate using 
the two tracers. Our work is supported by a large suite of mock catalogs, which we generated 
 by  performing 
 abbreviated N-body methods \citep[COLA,][]{tassev_etal:13}
to model potential systematics coming from observational issues, test different RSD models  
and their regime of validity, and determine covariances.

A potential advantage of a multi-tracer analysis was described by 
  \cite{mcdonald_seljak:09}, who noted that the correlations in an overlapping volume, if 
  the number density of the tracers is large, can be used to reduce the sample variance error
  and improve the measurements of the growth rate. 
  After this initial work, different applications of the multitracer method
  have been explored by various authors, using
 different observables such as  photometric redshift surveys, weak lensing, 
 gravitational redshifts, signatures of first 
 stars and constraints on primordial non-gaussianity and modified gravity 
\citep{seljak:09, bernstein_cai:11, gaztanaga_etal:12,asorey_etal:13,croft:13, yoo_etal:13, lombriser_etal:13}.
\cite{blake_etal:13} applied this method to the GAMA survey, producing modest gains from the
multitracer method, up to 20\% in the constraints of $f$ at two different epochs, $z=0.18$ and $z=0.38$.
\cite{aross_etal:14} measured the clustering of BOSS galaxies as a function of their colour and did not 
detect significant differences in distance scale or structure growth measurements. 
 Although the datasets  used in our study are too sparse to expect large improvement, 
 we include this effect by computing the full covariance of the measurements using our mock 
 galaxy catalogs.

We present in section  $\S$\ref{sec:data} the surveys used in our study.
In $\S$3 we present the methods and results of the auto- and cross- correlation between tracers. 
In $\S$4 we show models of the RSD and constraints in the model parameters and the growth rate 
at $z=0.54$. Finally in $\S$5 we summarize our results and conclude. This is the second
work of a series of papers analysing clustering in the BOSS-WiggleZ overlap region. Paper I 
(Beutler et al., 2015) focuses on the analysis of the  Baryonic Acoustic Oscillation signal 
of these two tracers in the common volume. 

For clarity we will use the name `CMASS-BW' and
`WiggleZ-BW' for the CMASS and WiggleZ samples limited
to the overlap region between the two surveys.
We assume a fiducial flat $\Lambda$CDM  cosmological model as defined in 
\cite{komatsu_etal:09}, where the matter density is 
 $\Omega_{\rm{m}} = 0.273$, baryon density of $\Omega_{\rm{b}}$ = 0.045, 
 a spectral index of $n_s$ = 0.963, 
 an r.m.s. of density fluctuations averaged in spheres of radii at 8 $h^{-1}$Mpc of $\sigma_8$ = 0.81 and $h = 0.71$.
 The Hubble rate at redshift $z$=0 is 
 $H_0$= 100$h$ km s$^{-1}$ Mpc$^{-1}$ is adopted 
 to convert redshifts to distances, which are  measured in $h^{-1}$Mpc.

\section{Data \& Mock catalogs}
\label{sec:data}

\subsection{The WiggleZ survey}

The WiggleZ Dark Energy Survey \citep{drinkwater_etal:10} is a large-scale galaxy redshift survey performed over 
276 nights with the AAOmega spectrograph \citep{sharp_etal:06} on the 3.9m Anglo-Australian Telescope. 
With a area coverage of 816 deg$^2$, this survey has mapped $207,000$ bright emission-line galaxies over a redshift range
$0.1<z<1.0$.   Target galaxies in six different regions  were chosen using
UV photometric data from the \emph{GALEX} survey \citep{martin_etal:05}
 matched with optical photometry from the
Sloan Digital Sky Survey (SDSS DR4,  \citealt{adelman_etal:06}) and 
from the Red-Sequence Cluster Survey 2 (RCS2, \citealt{gilbank_etal:11}).
The selection criteria consisted of applying magnitude and colour cuts \citep{drinkwater_etal:10} in order to select
 star-forming galaxies with bright emission lines with a redshift distribution centered around $z\sim0.6$.
 The selected galaxies were observed in 1-hour exposures using the 
 AAOmega spectrograph, and their redshifts were estimated from strong emission lines.  The number density of 
 WiggleZ galaxies averages $\sim 3 \times 10^{-4}$ $(h^{-1}$Mpc)$^{-3}$ at $z=0.6$.

\subsection{The CMASS Sample}
The Baryon Oscillation Spectroscopic Survey of the Sloan Digital Sky Survey III 
\cite[SDSS-III,][]{eisenstein_etal:11,dawson_etal:13}, which is now complete, 
was designed to obtain spectra and redshifts for 1.35 million bright  galaxies
over a footprint $\sim$ 10,000 deg$^2$. These galaxies are selected from the SDSS-III imaging and have been
 observed together with 160,000 quasars and 100,000 ancillary targets
 \citep{gunn_etal:06, bolton_etal:12, smee_etal:13}.
The CMASS sample is composed of luminous, mostly 
red galaxies selected
to probe large-scale structure at intermediate redshifts, achieving
 a number density of $\sim 3 \times 10^{-4}$ $(h^{-1}$Mpc)$^{-3}$.
The DR11 catalog \citep{alam_etal:15} includes 1,100,000 spectra out of  which the CMASS sample contains
 $\approx$ 550,000 galaxies in the redshift range $0.43 < z < 0.7$.

 \begin{table*}
\centering
 \caption{\label{tab:samples}Overlapping samples analysed in this study.}
  \begin{tabular}{l c c c c c}
  \hline
   Region & WiggleZ-BW & WiggleZ-BW & CMASS-BW& CMASS-BW & Cross-pairs\\
& $N_{gal}$ & $z_{avg}$ &  $N_{gal}$ & $z_{avg}$ & $z_{avg}$\\
 \hline
 S01& 6620  &0.61 & 5720 & 0.53  & 0.57\\
 N09& 13940 &0.56 & 9360  & 0.53  & 0.54\\
 N11& 15560 &0.55 & 10580 & 0.53  & 0.54\\
 N15& 22740 &0.56 & 14660 & 0.54  & 0.54\\
 S22& 10320 &0.55 & 6060  & 0.53  & 0.54\\
  \hline
Total & 69180 & 0.56 & 46380 &0.53 & 0.54\\
\hline
\end{tabular}
\end{table*}

\subsection{Overlap volumes}

We define the overlap regions between CMASS and WiggleZ using the 
random galaxy catalogs generated for each survey,
gridding the
sky into 0.1 deg$^2$ regions and selecting cells containing both CMASS and WiggleZ
random points.
As seen in Figure \ref{fig:allov}, five of the
six WiggleZ regions have considerable overlap with CMASS galaxies, totalling 560 deg$^2$ and 
a  volume of 0.218 ($h^{-1}$Gpc)$^3$ in the $0.43<z<0.7$ range. This results in an overlap 
sample of 69,180 WiggleZ galaxies and 46,380 CMASS galaxies. 

\begin{figure}
\includegraphics[width=8cm]{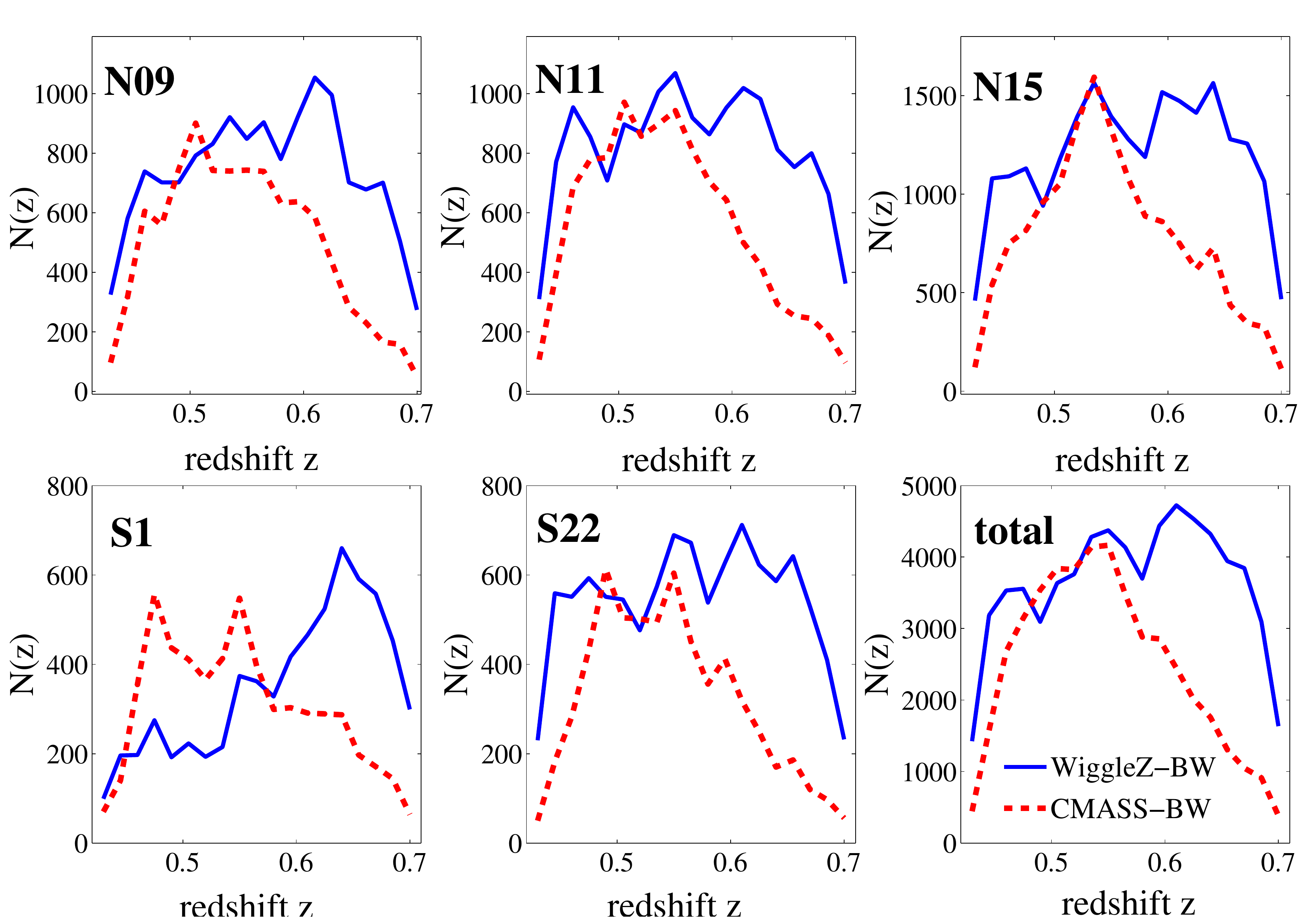}
\caption{\label{fig:redz_overlap}
Number of galaxies as a function of redshift for WiggleZ (blue) and CMASS (red) galaxies in the 
overlap regions, and for the overall overlap volume (bottom right panel).}
\end{figure}

Figure \ref{fig:redz_overlap} shows the redshift distribution of the two samples in the different regions, which 
is similar in the range  $0.5<z<0.6$; outside that range the CMASS galaxy counts rapidly decline. 
To estimate how these differences will affect our results,
we calculate the pair-weighted redshift,
which consists in taking the average redshift of all pairs at a particular distance range.
For the distance range $s=20-35$ $h^{-1}$Mpc, where the signal of the clustering 
signal is higher, 
WiggleZ-BW galaxies have a pair weighted redshift of $z_{\rm WiggleZ-BW}= 0.56 $ 
whereas for CMASS-BW galaxies  $z_{\rm CMASS-BW}= 0.53$.  
For cross-pairs this redshift is $z_{avg,\times}=0.54$.
These small differences in redshift  will not affect our 
findings given the measurement errors, 
therefore we generate cosmological models at $z_{avg,\times}=0.54$ to compare with
our WiggleZ-CMASS clustering data. Table \ref{tab:samples}
presents details of the samples used.

\subsection{Simulations and mock catalogs}

We estimate the covariance of our measurements and test the regime of validity of
our RSD models using  mock galaxy catalogs built from N-body simulations. The conventional
methods to generate N-body simulations do not allow for the generation of
 a large number of realisations of cosmological volumes at sufficient mass resolution 
 to encompass the low-mass halos hosting WiggleZ galaxies,  
 which are needed for constructing robust covariance matrices. For this
reason we use an approximate, fast method to generate dark matter simulations based on 
the COmoving Lagrangian Acceleration method \citep*[COLA,][]{tassev_etal:13}.
We have developed a parallel version of COLA \citep[Koda et al., in preparation, used first in][]{kazin_etal:14}, where in each simulation contains 
$1296^3$ particles in a box of side
$600h^{-1}\mathrm{Mpc}$, which gives a particle mass of
$7.5\times10^{9}h^{-1}M_\odot$, allowing resolution of   low-biased halos 
with masses  $10^{12}$ $h^{-1}M_\odot$,
found using friends-of-friends algorithm
 with a linking length of $0.2$ times the mean particle separation.
 Each simulation requires 15 minutes with 216 computation
cores, including halo finding, which is much faster than a classical N-body simulation, but
with similar precision on the relevant scales ($k<1$ $h$Mpc$^{-1}$).

\begin{figure}
\includegraphics[width=8.5cm]{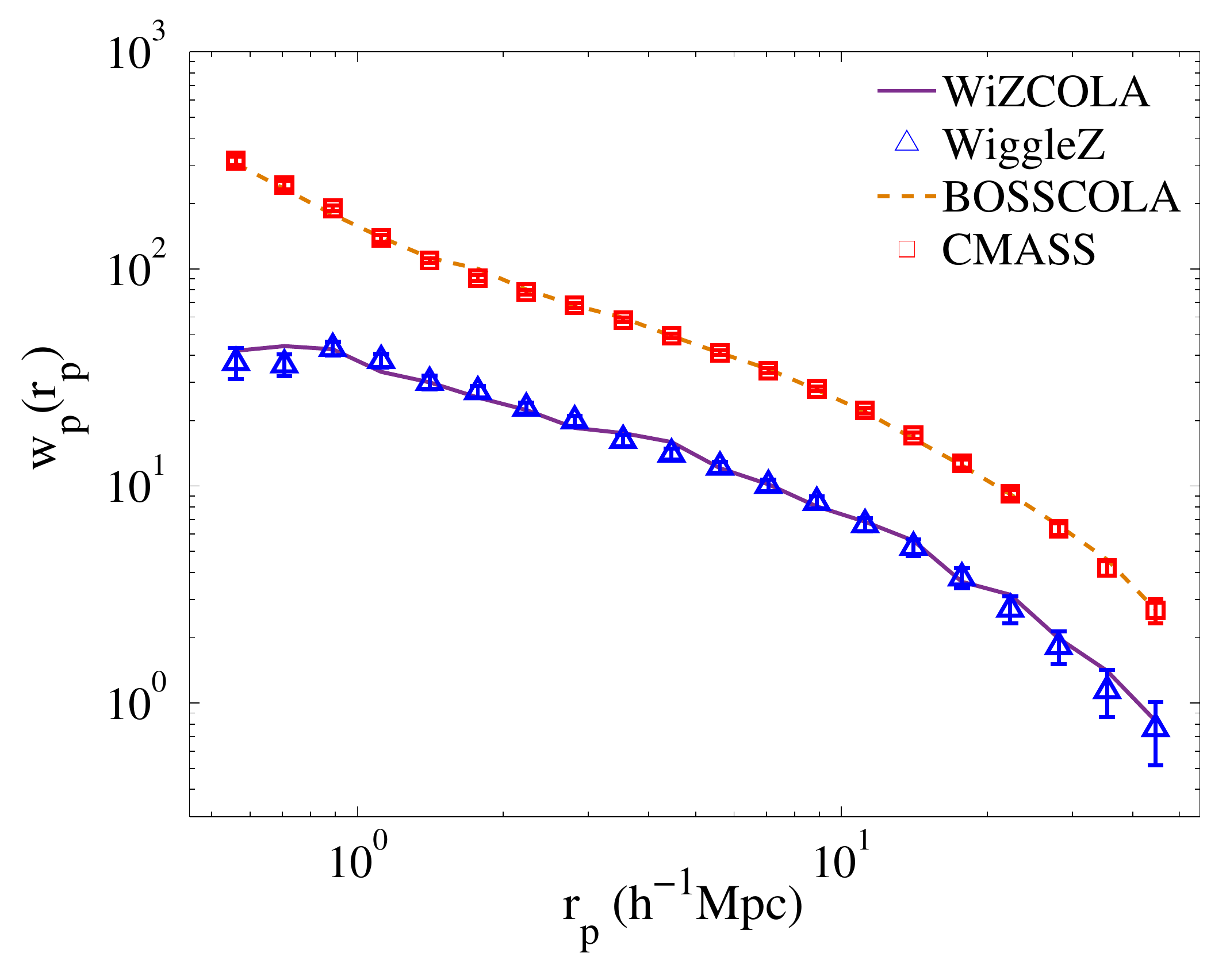}
\caption{\label{fig:wprp}
Projected correlation function $w_p(r_p)$ for WiggleZ  and CMASS  galaxies in the 
overlap regions (symbols). Lines are the mean values of $w_p(r_p)$ for the
COLA mock catalogs.}
\end{figure}

We generate a total of 2400 realisations (480 for each WiggleZ region) of a flat  $\Lambda$CDM 
universe with WMAP5 cosmological parameters (Komatsu et al., 2009), which defines our fiducial
cosmology. 
 Using the 
output at $z=0.6$ we create WiggleZ-based (WiZcola) and CMASS-based (BOSScola)
mock galaxy catalogs, from simple Halo Occupation Distribution models \citep{berlind_weiberg:02, blake_etal:08}, 
such that the resulting projected correlation 
functions $w_{\rm p}(r_{\rm p})$ match those of the observations, as seen in Figure 
\ref{fig:wprp}. We then apply the relevant selection functions to the mock galaxies to
match the  survey geometry. Our simulations encode the joint  covariance in the overlapping survey regions (Koda et al., in prep.).

\section{Measurements}

\subsection{Measuring Correlation Functions}

\begin{figure*}
\includegraphics[width=16cm]{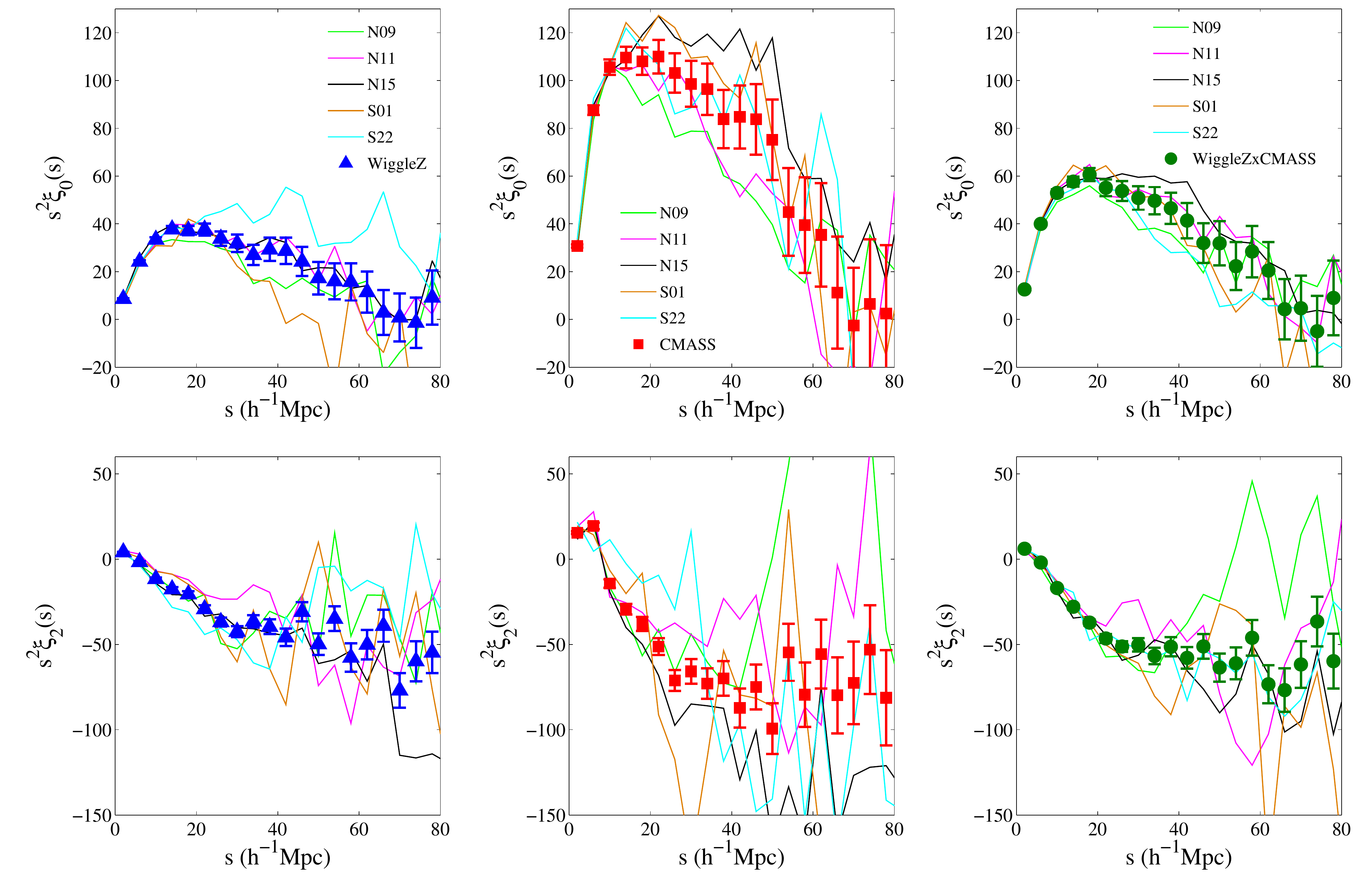}
\caption{\label{fig:monoquad}
Combined monopole (top row) and quadrupole (bottom row) of $\xi(s,\mu)$ for the WiggleZ and CMASS auto-correlation function (left and middle columns) and the cross-correlation function (right column).  Lines are measurements of individual regions, symbols display the combined measurements. Results are plotted as $s^2\xi_l(s)$ as a function of separation $s$.}
\end{figure*}

We estimate the redshift-space two-point correlation function $\xi(s,\mu)$  (2PCF) 
as a function of comoving separation $s$ and 
the cosine of the angle of the distance vector with respect to the line of sight $\mu = \cos(\theta)$.
We use the \cite{landy_szalay:92} estimator, counting pairs of objects in data and
random catalogs: 
\begin{equation}
\xi_{\rm auto}(s,\mu)= \frac{DD(s,\mu) - 2DR(s,\mu) + RR(s,\mu)}{RR(s,\mu)},
\end{equation}
where  $DD$, $DR$ and $RR$ are respectively the weight-normalised data-data, data-random and 
random-random pairs with separation $s$ and $\mu$ (with a given resolution $\Delta s$, $\Delta \mu$).
For both the random and data catalogs we use the optimal (inverse-density) FKP weighting \citep{feldman_etal:94}: 
\begin{equation}
w_i({\bf x}) =  \frac{1}{1+n({\bf x})P_0}
\end{equation}
where $P_0=5000$ $(h^{-1}$Mpc)$^{3}$ for WiggleZ-BW and $P_0=20000$ $(h^{-1}$Mpc)$^{3}$ for CMASS-BW galaxies.
For WiggleZ galaxies, angular incompleteness and radial selection are introduced in the random catalogs 
\citep{blake_etal:10sf}. A small fraction of galaxies contain errors in the redshift assignment, but this effect is absorbed
into the fitted galaxy bias factor. 
CMASS galaxies, have additional weights applied to account for 
the angular incompleteness, fibre collisions, redshift failure and correlation
between density of targets and density of stars \citep{rossaj_etal:12}. 

It is possible to model the 2PCF using the full information from $\xi(s,\mu)$, but that requires a
large covariance matrix with the associated problems with its inversion. 
For this reason it is standard to compress this information in multipoles
\begin{equation}\label{eq:multil}
\xi_l (s)= \frac{2l+1}{2}\int_{-1}^{1}\xi(s,\mu)L_l(\mu)d\mu,
\end{equation}
where $L_l$ is the Legendre polynomial of order $l$. In practice we approximate eq. (\ref{eq:multil}) by a discrete
sum over the binned $\xi(s,\mu)$, where we use $\Delta s=4$ $h^{-1}$Mpc and $\Delta \mu$=0.01 for every 
WiggleZ-BW and CMASS-BW region. We use the monopole ($l=0$)  and quadrupole ($l=2$)
of the 2-point functions, to analyse the redshift-space distortions, for separations $s<80$ $h^{-1}$Mpc.
Our results are unchanged if large separations are used, whilst the increase in variance due to the 
finite number of mocks becomes significant.

The covariance of each region is estimated from the mock WiZcola and BOSScola catalogs (see section 3.3).
After calculating the covariances of the measurements in each overlap region from the COLA mock catalogs,
we use
inverse-variance  weighting to obtain the `optimally combined' measurements.
For  the statistic $\xi_{l,\rm{comb}}(s)$, the optimally combined function is calculated as 
\begin{equation}\label{eq:xcomb}
\xi_{l,\rm{comb}}(s) = \mathcal{C}_{\rm{comb}} \sum_{i=1}^{N_{reg}} \mathcal{C} _i^{-1}\xi_{l,i}(s)
\end{equation}
where  $\mathcal{C}_{\rm{comb}}$ is the overall covariance matrix, calculated from the estimations of the 
covariance matrices of individual regions $\mathcal{C} _i$ (see section 3.3).
Results for the auto-2PCFs
 are shown in Figure \ref{fig:monoquad}, for individual regions (as lines) and for the combined measurements 
 (as symbols). 
The different  amplitude of clustering of the WiggleZ-BW and CMASS-BW galaxies reflects  the difference in the type of halos these 
galaxies inhabit.
Due
to the limited volume where the correlations are measured, 
we correct our correlation function values by the `integral constraint' \citep{peebles:80, beutler_etal:12rsd}.
 The corrections to
the WiggleZ and BOSS correlations differ 
in each region and have values of 
 the order of $8\times10^{-4}$ and $1\times10^{-3}$ respectively for the 
smaller regions (where the integral constraint is higher), and do not  significantly affect 
the RSD model constraints.

\begin{figure}
\includegraphics[width=8cm]{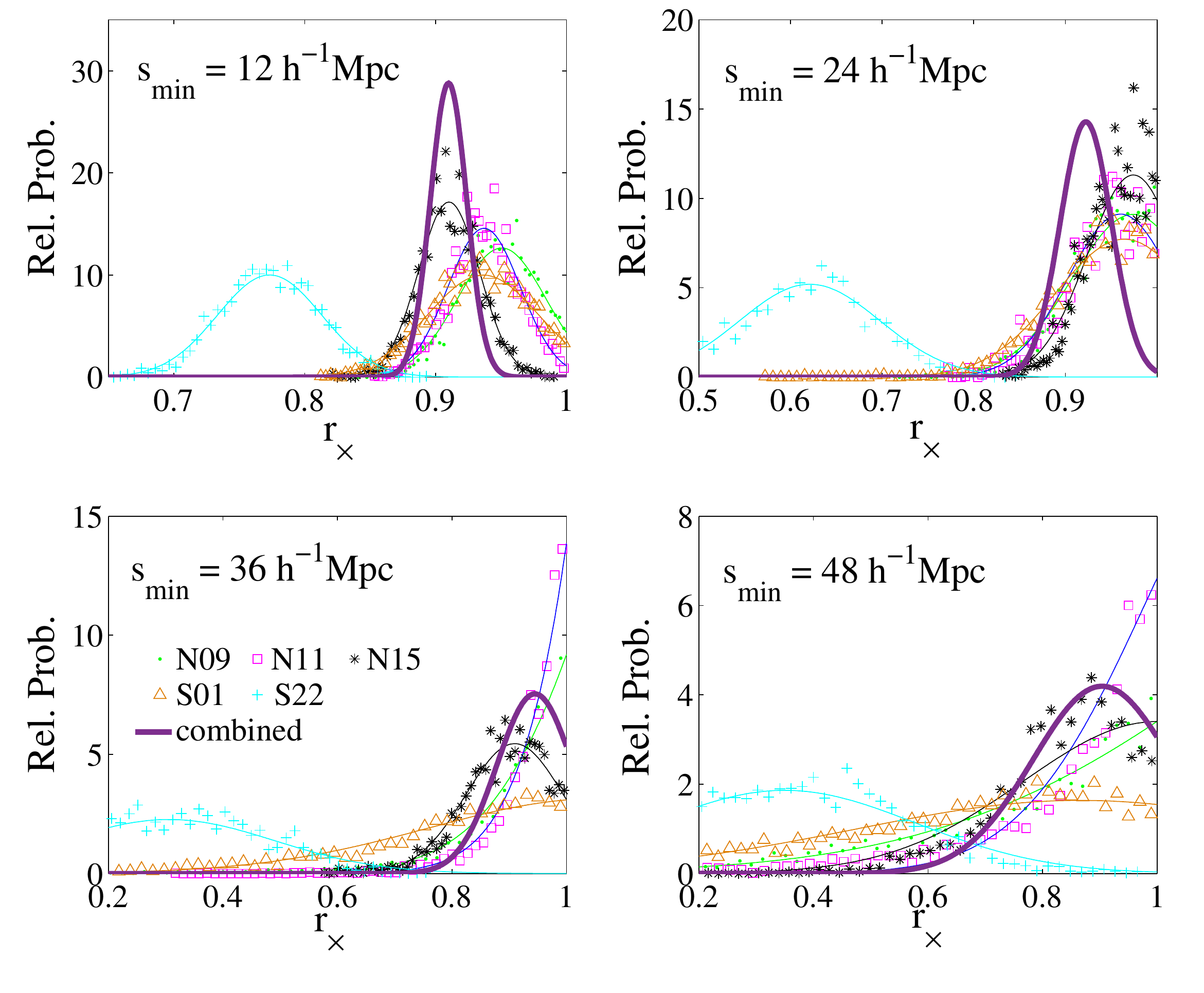}
\caption{\label{fig:rcross}
The cross-correlation coefficient $r_\times$ for the WiggleZ-CMASS-BW correlations for each overlap regions, and when 
combining all regions, as a function of the smallest  scale $s_{min}$ (in $h^{-1}$Mpc) of the fit.}
\end{figure}

\subsection{Cross-correlations between WiggleZ-BW and CMASS-BW clustering}

In addition to the auto-correlations, we also measured the cross-correlation
between the two sets for tracers using the estimator
\begin{equation}
\xi_{\rm cross}(s,\mu)= \frac{D_WD_C - D_WR_C - R_WD_C+ R_WR_C}{R_WR_C},
\end{equation}
where the $W$ and $C$ subscripts represent the quantities in the WiggleZ and CMASS galaxies, respectively. 
The cross-correlation function measurement provides an independent validation of the assumption that both
galaxy types trace the same large structures on a range of scales, and also serves to test our linear and local galaxy bias model.

To test the strength of the correlation between the tracers is, we also constrain the  cross-correlation
coefficient, $r_\times$, which is produced from the relation
\begin{equation}
\xi^{l=0}_{\rm WiggleZ\times CMASS}(s) =   r^2_\times(s) \xi^{l=0}_{\rm WiggleZ}(s)\xi^{l=0}_{\rm CMASS}(s),
\end{equation}
with $|r_\times|\leq 1$.
On large scales in redshift-space, and assuming linear, deterministic bias,  this quantity should tend to the value
\citep{mountrichas_etal:09}:
\begin{equation}
r_{\times,{\rm Kaiser}} = \frac{1+\frac{1}{3}(\beta_W+\beta_C)+\frac{1}{5}\beta_W\beta_C}
{\sqrt{\left(1+\frac{2}{3}\beta_W+\frac{1}{5}\beta_W^2\right)\left(1+\frac{2}{3}\beta_C+\frac{1}{5}\beta_C^2\right)}}
\end{equation}
Assuming $b_W=1$ and $b_C=2$  \citep{reid_etal:12,contreras_etal:13}, and a 
growth rate $f(z=0.54)=0.75$, when estimating  $\beta=f/b$,
it is expected that 
$r_{\times,{\rm Kaiser}} = 0.997\sim 1$.  

We  measure the value of $r_\times$ from our data, assuming it is a constant on all scales (an assumption we do not 
expect to hold on scales smaller than $~ 15-20$ $h^{-1}$Mpc).
Using the redshift space distortion model described in section 4.1, 
and the COLA mocks to build our covariance matrix, we use the correlation monopoles to
fit for the bias parameters of the WiggleZ-BW and CMASS-BW galaxies and $r_\times$ for each overlap region and the
joint likelihood (see section 4.3 for details of the fitting procedure). 

 Figure \ref{fig:rcross} presents the posterior probability distribution of $r_\times$,
   as a function of the minimum scale of fit $s_{min}$. 
   Focusing on the fits to the combined regions, 
   we can see that they are not consistent with $r_\times=1$ at the 
 $2\sigma$ level on scales $s_{min}\sim20$ $h^{-1}$Mpc.
 This behaviour may be explained by a number of factors such as non-linear pairwise velocities, 
 non-linear bias and stochasticity.  CMASS galaxies tend to be hosted in 
 the centres of large halos and in high density regions, precisely the regions that are avoided by WiggleZ galaxies.
  We expect
 that on large scales both galaxies trace similar structures,
  and this is confirmed in the measurements of $r_\times$ being consistent
 with 1 when fitting on large scales.
 
 Examining individual regions it can be noticed that it is region S22 which reduces
 the overall fit to $r_\times$. Its lower value of $r_\times$ is driven by a high auto-correlation function 
 in  the WiggleZ-BW S22 region, although the scatter is compatible with the variance against mock catalogs. 
The best fits to the growth rate do not significantly change when the
S22 region is excluded, and in the final fits we include all regions.

\subsection{Covariance estimation}

\begin{figure}
\includegraphics[width=8.5cm]{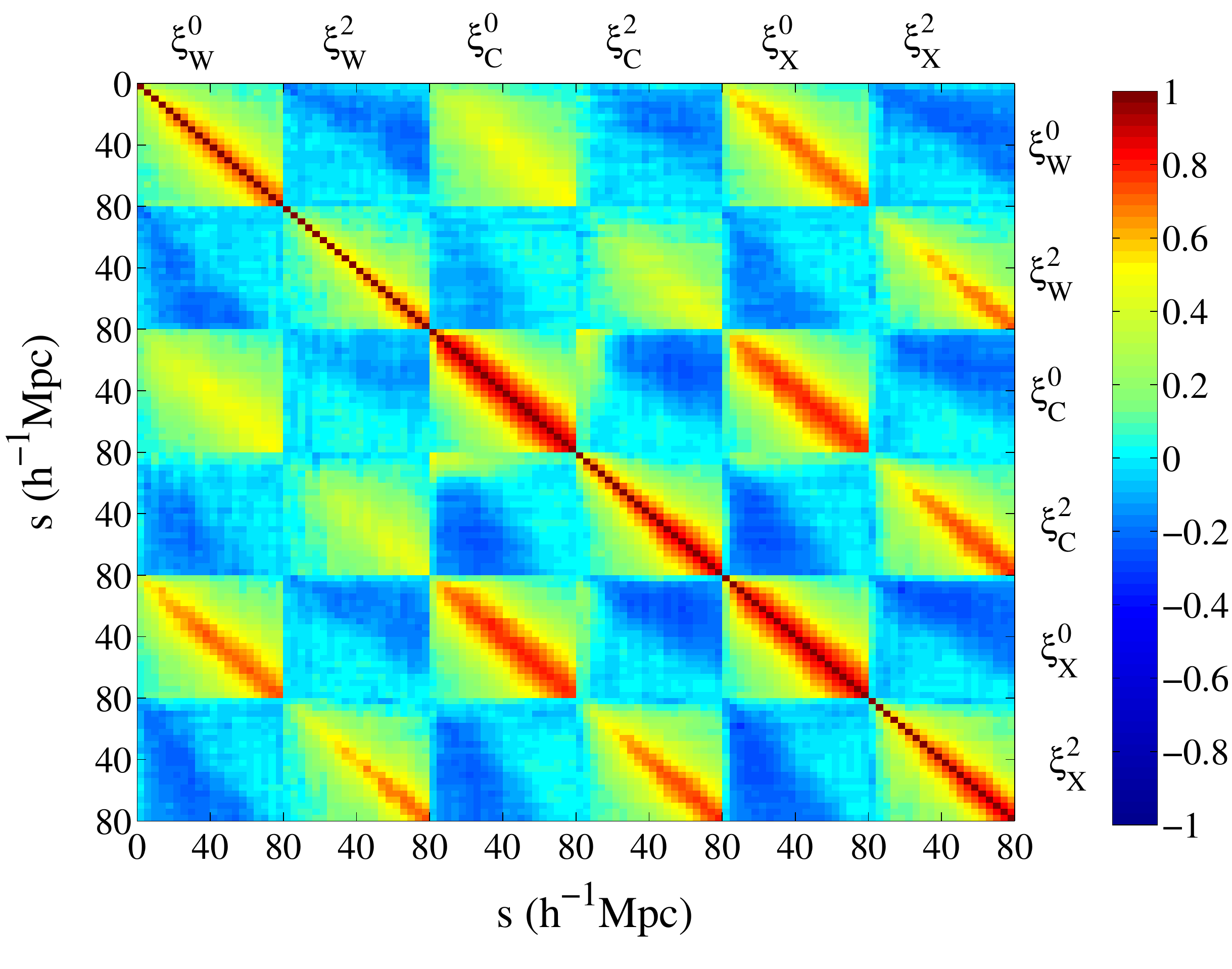}
\caption{\label{fig:corrmatrix}
Correlation matrix (normalised covariance matrix) of the WiggleZ-CMASS multipoles based on COLA mocks.
The matrix is divided in the contributions from the monopole and quadrupole of WiggleZ-BW (W), CMASS-BW (C) and 
cross 2PCFS (X), and each pixel represents a separation s (in $h^{-1}$Mpc).}
\end{figure}

We  estimate the correlations between the multipoles of the auto- and cross-2PCF by
 calculating the covariance matrix in each region $n$ from  COLA mocks.  
 A deviation from the mean of a quantity $X$, in separation bin $i$, for the mock $k$ can be written as
 \begin{equation}
 \Delta_{i,n}^k = X_i^k-\langle X_i \rangle
 \end{equation}
 where, in our case, $X$ corresponds to the monopole or quadrupole of the auto- or cross-2PCF in each bin.
The  covariance matrix of each region $n$  is determined as 
 \begin{equation}
 \mathcal{C}_{n,ij}=\frac{1}{N_{\rm mocks}}\sum_{k=1}^{N_{\rm mocks}} \Delta_{i,n}^k \Delta_{j,n}^k
 \end{equation}
After calculating
$ \mathcal{C}_n$ for all regions, we can determine the combined covariance matrix \citep{blake_etat:11bao}
\begin{equation}
\mathcal{C}^{-1}_{comb}=\sum_{n=1}^{N_{reg}}\mathcal{C}_n^{-1}
\end{equation}
Figure \ref{fig:corrmatrix} shows the correlation matrix (normalised covariance matrix) for all our measurements, 
showing the strong correlation between the measurements of the two tracers. 

Since we used a large, but finite number of mock catalogs for the covariance estimation, there  is an underestimation
of the uncertainties. Following the work of  \cite{hartlap_etal:07,percival_etal:14}, we correct for the finite number
of mocks by multiplying  the variance estimated from the likelihood distribution by
\begin{equation}
m_\sigma=\frac{1 + B(N_{\rm bins}-N_{\rm p})}{1+2A+N(N_{\rm p}+1)}
\end{equation} 
where $N_{\rm bins}$ is the number of bins entering the fits, $N_{\rm p}$ is the number of free parameters, and
\begin{eqnarray}
A^{-1} &=& (N_{\rm mocks} - N_{\rm bins} - 1)(N_{\rm mocks} - N_{\rm bins} -4),\\
B &=& A(N_{\rm mocks} - N_{\rm bins} - 2),
\end{eqnarray}
where $N_{\rm mocks}$ is the number of mock realisations. Also, the sample variance should be multiplied by 
\begin{equation}
m_v = m_\sigma\frac{N_{\rm mocks} -1}{N_{\rm mocks} - N_{\rm bins} - 2}.
\end{equation}
We use $N_{\rm mocks}= 480$ and perform measurements in separation bins up to $s=80$ $h^{-1}$Mpc. From constraining models using  one-tracer
auto-correlation function multipoles to simultaneous fits using both auto and cross correlations, 
the $m_v$ factor lies in the range 
$m_v=1.1-1.45$.

\section{Constraints on cosmic growth}

\subsection{Modelling the RSD}

Redshift-space distortions modify  the 2-point clustering of galaxies on both large and small scales, 
which we will summarise here. Due to its peculiar velocity ${\bf v}$, a galaxy at a position in real space ${\bf r}$
gets mapped to ${\bf s}$ in redshift space:
\begin{equation}
{\bf s} = {\bf r} + \frac{(1+z){\bf v} \cdot \hat{r}}{H(z)}\hat{r}
\end{equation}
where $\hat{r}$ is the galaxy unit vector along the line of sight (LOS)  direction, $v_r\equiv {\bf v} \cdot \hat{r}$ 
is the line-of-sight component of its velocity, and
$H(z)$  is the Hubble parameter at a redshift $z$.

On large scales, as described by \cite{kaiser:87}, hereafter K87 
(also see \citealt{hamilton:98} for derivations in configuration space), 
matter  overdensities $\delta_m$ grow coherently 
as $\nabla\cdot{\bf v} \propto -f\delta_m$ where $f\equiv d\ln G(a)/d\ln a$ is 
the linear growth rate of fluctuations. 
The evolution of the growth rate in certain models can be approximated by
the evolution of the matter density in the universe $f(z)=\Omega_m(z)^\gamma$, where $\gamma=0.55$ 
in the case that  the large-scale gravity obeys General Relativity; 
for alternative theories of gravity, $\gamma$ can take on different
values \citep{linder_cahn:07}.
  If we assume that the difference in clustering between dark matter and galaxies can be described by a 
 linear bias model where $\delta_g = b \, \delta_m$ then in Fourier space the redshift space galaxy overdensity 
takes the form
\begin{equation}
\delta_g({\bf k}) = (b + f\mu^2)\delta_m({\bf k}),
\end{equation}
creating in configuration space the so-called `squashing' effect on large scales.

On small scales, in the non-linear regime for overdensities and velocities, 
large structures appear elongated
along the line of sight, 
creating the observed `Fingers of God'. In Fourier space this effect can be modelled by
 multiplying a Gaussian or a Lorentzian 
pairwise velocity distribution (i.e. a convolution of a Gaussian or exponential profile in configuration space) into the
large-scale redshift-space distortion of the power spectrum. 
The simplest  model, using Gaussian damping 
for the galaxy power spectrum in redshift space, is 
\begin{equation}
P^s(k,\mu)=(b+f\mu^2)^2P_m(k)e^{-(k\mu\sigma_v/H)^2}
\end{equation}
$P_m(k)$ represents the non-linear real-space power spectrum and $\sigma_v$ the pairwise 
velocity dispersion, which we approximate to be the same for all tracers is predicted 
to be 
\begin{equation}
\sigma_v^2(z) = \frac{f^2(z)H^2(z)}{6\pi^2(1+z)^2}\int P_{\theta\theta}(k)dk
\label{eqsigv}
\end{equation}
where, in the K87 formalism, $P_{\theta\theta}=P_m$.
However, this simple model has been shown in simulations to be insufficiently 
accurate 
even on large scales, because there is not a perfect correlation between
density and the velocity (divergence) field 
\citep[e.g][]{okumura_jing:11, kwan_lewis:11,delatorre_guzzo:12, white_etal:14}.
\cite{scoccimarro:04}, hereafter S04,  suggested a modification of the simple 
Kaiser formalism by including  the velocity field terms.
In the case of one tracer, the  RSD in the galaxy auto- power spectrum reads:
\begin{equation}
P^s_{\rm a}(k,\mu) = [b^2P_{\delta\delta}(k)+2\mu^2fbP_{\delta\theta}(k)+\mu^4f^2P_{\theta\theta}(k)]e^{-(k\mu\sigma_v/H)^2}
\end{equation}
where $P_{\delta\delta}$, $P_{\delta\theta}$ and $P_{\theta\theta}$ 
are the non-linear matter density-density, density-velocity 
and velocity-velocity power spectra, respectively. 
In our analysis, these terms are obtained from  fitting formulae derived by 
\cite{jennings:12},  from a suite of  N-body simulations.
In this case our fiducial model (based on WMAP5 results, see $\S$1) predicts (via Eq.\ref{eqsigv}) the large-scale 
velocity dispersion $\sigma_v$ to be
$\sigma_v(z=0.6)\sim 220$ kms$^{-1}$. However, we choose to leave $\sigma_v$ as a free parameter to account
for any additional non-linearities on smaller scales.
Whilst there are additional
  improvements and implementations  of  RSD models \citep[e.g.][]{taruya_etal:10, seljak_mcdonald:11, reid_white:11, wangl_etal:13},
this particular formalism has been successfully used in a number of studies
 \citep[e.g.][]{blake_etal:11f,delatorre_etal:13,blake_etal:13}, and, as we will see below, reproduces the expected constraints on 
 the growth rate from the COLA mock catalogs and provides a  good description of the galaxy  anisotropic
 clustering at the current statistical level. 

In the case of the redshift-space cross-power spectrum, 
assuming that both tracers are described by the same dispersion parameter $\sigma_v$, we
can write the large-scale terms as
\begin{eqnarray}
P^s_{\rm x}(k,\mu) &=& [b_1b_2P_{\delta\delta}(k)+\mu^2f(b_1+b_2)P_{\delta\theta}(k)+
\nonumber \\ 
&& \mu^4f^2P_{\theta\theta}(k)]\times e^{-(k\mu\sigma_v/H(z))^2}
\end{eqnarray}
where $b_1$ and $b_2$ are the biases of the different tracers.
Since we are measuring the multipoles of the 2PCF, we calculate first the power spectrum multipoles as
\begin{equation}
P^s_l(k)=  \frac{2l+1}{2}\int_{-1}^{1}P^s(k,\mu)L_l(\mu)d\mu,
\end{equation}
where $l$ is the multipole order and  $L_l$ is the Legendre polynomial of order $l$. Then for the 
2-point correlation function in configuration space we have
\begin{equation}
\xi^s_l(s) = \frac{i^l}{2\pi^2}\int P^s_l(k)j_l(ks)k^2dk
\end{equation}
where $j_l$ is the spherical Bessel function of order $l$.\\

\subsection{Tests using COLA mocks}

We tested the validity of these models using our COLA mock catalogs. 
In summary, we compared 
the K87 and S04 models for the large-scale distortions to $P(k)$ (calculated using our
fiducial cosmological parameters), 
using a Gaussian function for the small-scale damping (we also tried fits using the Lorentzian
profile without significant differences), constraining the growth rate $f$ at the simulation output redshift $z=0.6$, marginalising
over the bias of each tracer and the common velocity dispersion $\sigma_v$.
We performed these fits for every COLA mock on scales $s<80$ $h^{-1}$Mpc, 
although changes when using larger scales were not significant.

\begin{figure}
\includegraphics[width=8.5cm]{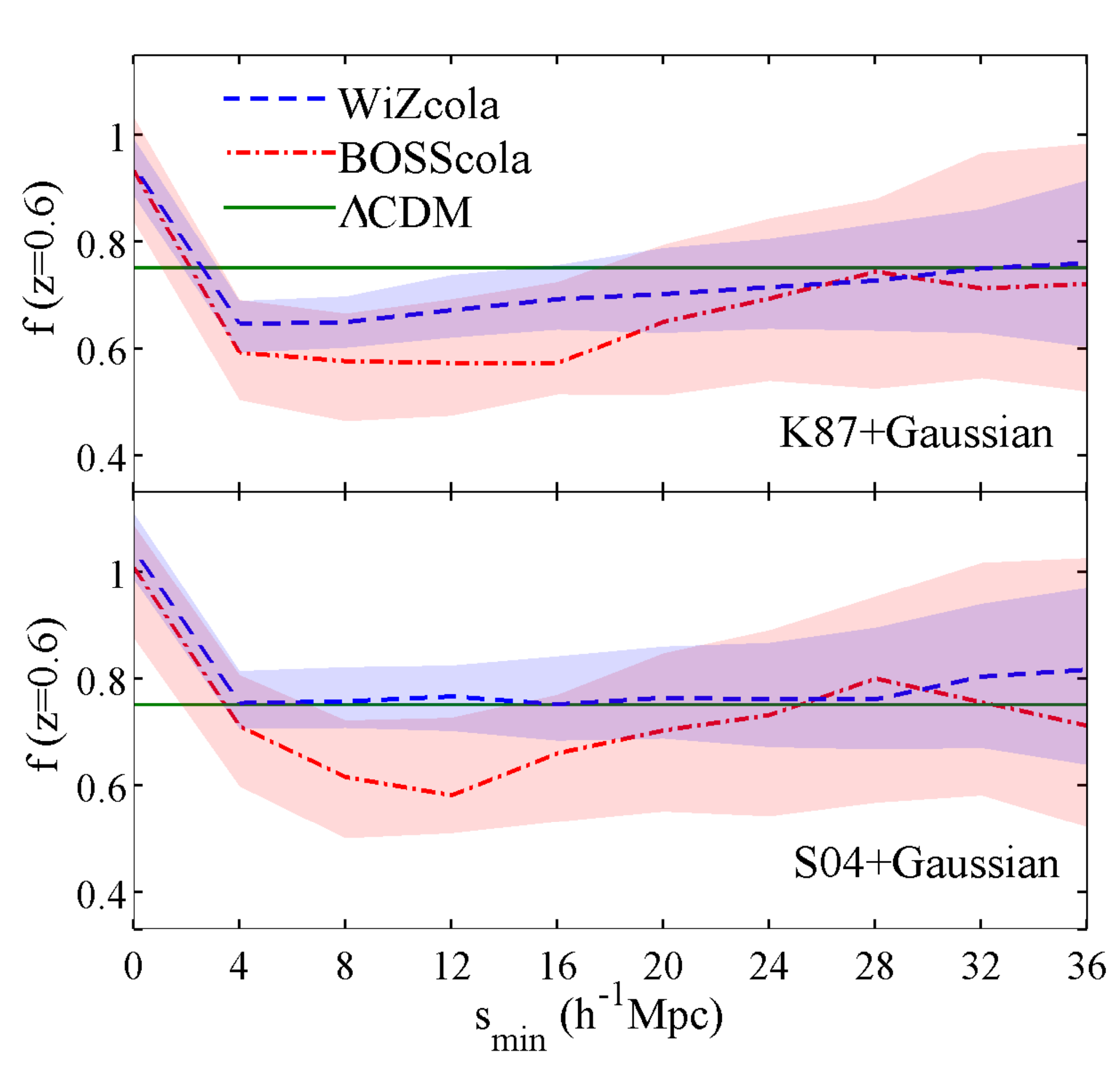}
\caption{\label{fig:f1autocola}
Constraints on the linear growth rate $f(z=0.6)$ using COLA mock galaxies with the same angular and redshift
selection functions as the galaxy survey data. Fits to the  growth rate $f(z=0.6)$ ($=0.76$, for 
the WMAP5 cosmology, green solid line)
are performed for the models as a function of $s_{min}$, with $s_{max}=80$
$h^{-1}$Mpc. Colored shades cover the 1$\sigma$ (68\%) confidence interval for $f$, 
defined using the dispersion in the best-fitting values across the 480 mocks.}
\end{figure}

Figure \ref{fig:f1autocola} shows the mean and standard deviation of the best-fitting values of $f(z=0.6)$ across the mocks.
For the WiZcola mocks 
 the K87+Gaussian model tends to underpredict the value
of the growth rate whereas the S04+Gaussian model agrees well for scales $s_{\rm min} > 20$ $h^{-1}$Mpc.
For the BOSScola mocks the differences are less pronounced, but the input growth rate is recovered with
a systematic error less than the statistical error.
In both cases the goodness of fit is similar with $\chi^2$/d.o.f.$\sim 1$ for both WiggleZ and BOSS COLA mocks
on larger scales
$s_{\rm min}>20$ $h^{-1}$Mpc, worsening considerably on scales $s_{\rm min}<10$ $h^{-1}$Mpc.
In what follows we will use the S04+Gaussian model for our parameter fits.

There are, however, specific
differences in the scale of validity of the models depending on which tracer is used. It can be seen
that for low-bias galaxies represented by the WiZcola mocks, the agreement between the model fits and 
the input value
of $f(z=0.6)$  extends to lower scales than in the case of galaxies residing in more massive halos. 
Although the Kaiser effect is stronger for lower bias galaxies, the higher non-linearities
arising from the formation and high-clustering of high-mass halos lead to a model break-down on larger scales. 
For the particular case of the WiggleZ-BOSS overlap, 
 $s_{min}=24$ $h^{-1}$Mpc is the minimum scale where both models recover adequately 
the fiducial growth rate with negligible systematic error.

\subsubsection*{Multitracer approach}

\begin{figure}
\includegraphics[width=8.5cm]{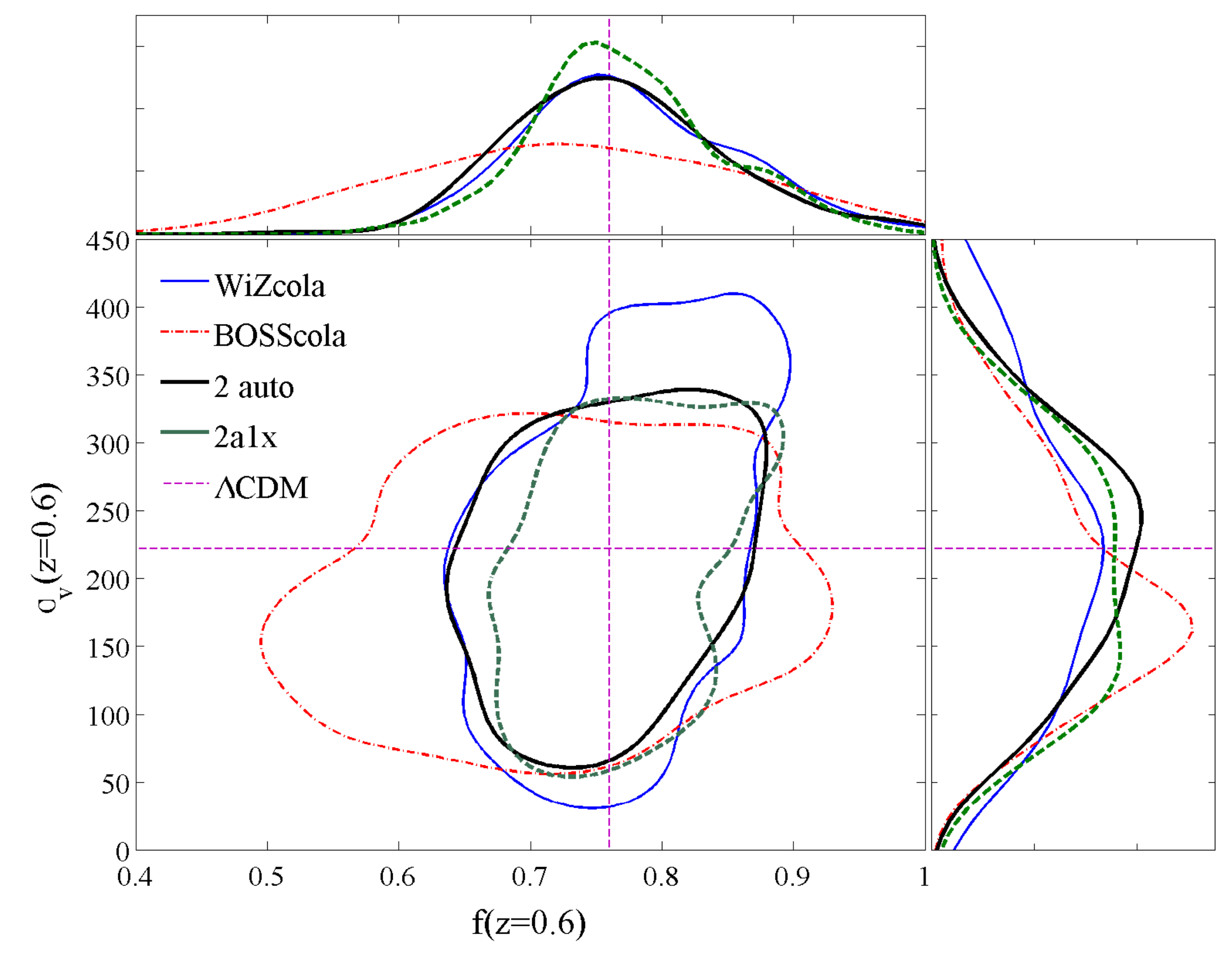}
\caption{\label{fig:colafits24}
 Constraints  for RSD parameters in the S04 model for fits to  COLA mocks in the range  $24<s<80$ $h^{-1}$Mpc using
 2PCF multipoles. Results for the full fit and for individual subsamples, with the contours enclosing 
 68\% of the best-fitting parameters are shown. The four cases considered are WiZcola mocks alone (blue solid line), 
 BOSScola mocks alone (red dashed line), joint analysis of the auto-correlations (black solid line),
  and an analysis further adding the cross-correlation function (green dashed line).}
 \end{figure}

Having chosen the  model to analyse redshift-space clustering,  we examined the consequences of using multiple tracers when recovering   model parameters. For each realisation of the COLA mocks
 we fit the S04+Gaussian RSD model first
using the autocorrelations independently, then analysing both autocorrelations but considering the common 
covariance matrix, and lastly adding the cross-correlations 
using the monopole and quadrupole
of $\xi(s,\mu)$ in the range $24<s<80$ $h^{-1}$Mpc. 
Results are shown in Figure \ref{fig:colafits24}, which displays
the 1$\sigma$ contours enclosing the best-fitting values of $f(z=0.6)$ and $\sigma_v$. The 
expected values for these parameters are consistent with our COLA constraints, and the approximation
that $\sigma_v$ is the same for both tracers is valid for this range of scales. 
 The constraints for the parameters are correlated between the two surveys, 
 with a cross-correlation
coefficient $\rho^f_{WC} = \sigma^f_{WC}/\sigma^f_{W}\sigma^f_{C} = 0.4$ for the growth rate between both surveys.

When analysing the 2PCFs of the two tracers simultaneously, taking into account
the common covariance, an improvement in the measurement of $f$  is obtained, of the order
30\% compared to using the BOSScola mocks alone (which because of a higher bias, have a lower value of
$\beta$ and hence a lower signal) but only 5\%  compared to using
WiZcola mocks alone. 
Adding the cross-2PCF  produces
an improvement of 20\%
compared to the WiZcola-only constraints, mostly due to an increased signal in the shot-noise dominated regime.
Analysing individual mocks shows that the improvement also varies in each realisation. 
As predicted by  \cite{mcdonald_seljak:09}, \cite{gil-marin_etal:10} and \cite{blake_etal:13}, 
although our tracers have big differences in their 
biasing, due to the sparsity of our sampling  we 
are in the regime where shot noise dominates and improvement via the cancellation of cosmic
variance is small.

\subsection{Data Fitting procedure}
In our analysis we fixed the cosmological parameters of the matter power spectra to the best-fit WMAP5 model
\citep{komatsu_etal:09}, the fiducial cosmology of our COLA mocks, and
constrain the parameters $(b_W,b_C,f(z=0.54), \sigma_v)$. Due to the degeneracy
of the first three  parameters with
$\sigma_8(z)$, the r.m.s. of the matter density field in 8 $h^{-1}$Mpc spheres,
we are effectively constraining 
$(b_W\sigma_8,b_C\sigma_8,f\sigma_8, \sigma_v)$.
When we also include the WiggleZ-CMASS cross-correlation in the analysis, we additionally fit
for the parameter $r_\times$. We  compare the constraints from the single-tracer model for
each galaxy type to each other, and then include the common covariance and the cross-correlations
in the cosmological fits.

We use the monopole
and quadrupole of the tracers, and present results as a function  of the minimum-scale fitted $s_{\rm min}$, 
with $s_{max}=80$ $h^{-1}$Mpc. We execute a Maximum Likelihood parameter estimation test, where 
we minimise the quantity
\begin{equation}
\chi^2=\sum_{i}\sum_{j}(X_{i,{\rm model}}-X_{i,{\rm data}})\mathcal{C}_{comb,ij}^{-1}(X_{j,{\rm model}}-X_{j,{\rm data}}),
\label{eq:chi2}
\end{equation}
where $X$ is one of the elements of the vector formed by the multipoles of $\xi(s,\mu)$ of WiggleZ-BW, 
CMASS-BW and/or WiggleZ-CMASS-BW correlations. 
We explore the parameter space using a Monte Carlo Markov 
 Chain method (MCMC) imposing the prior that all parameter values
 must be bigger than zero.

\subsection{Fits for the growth rate}

\begin{figure}
\includegraphics[width=8.5cm]{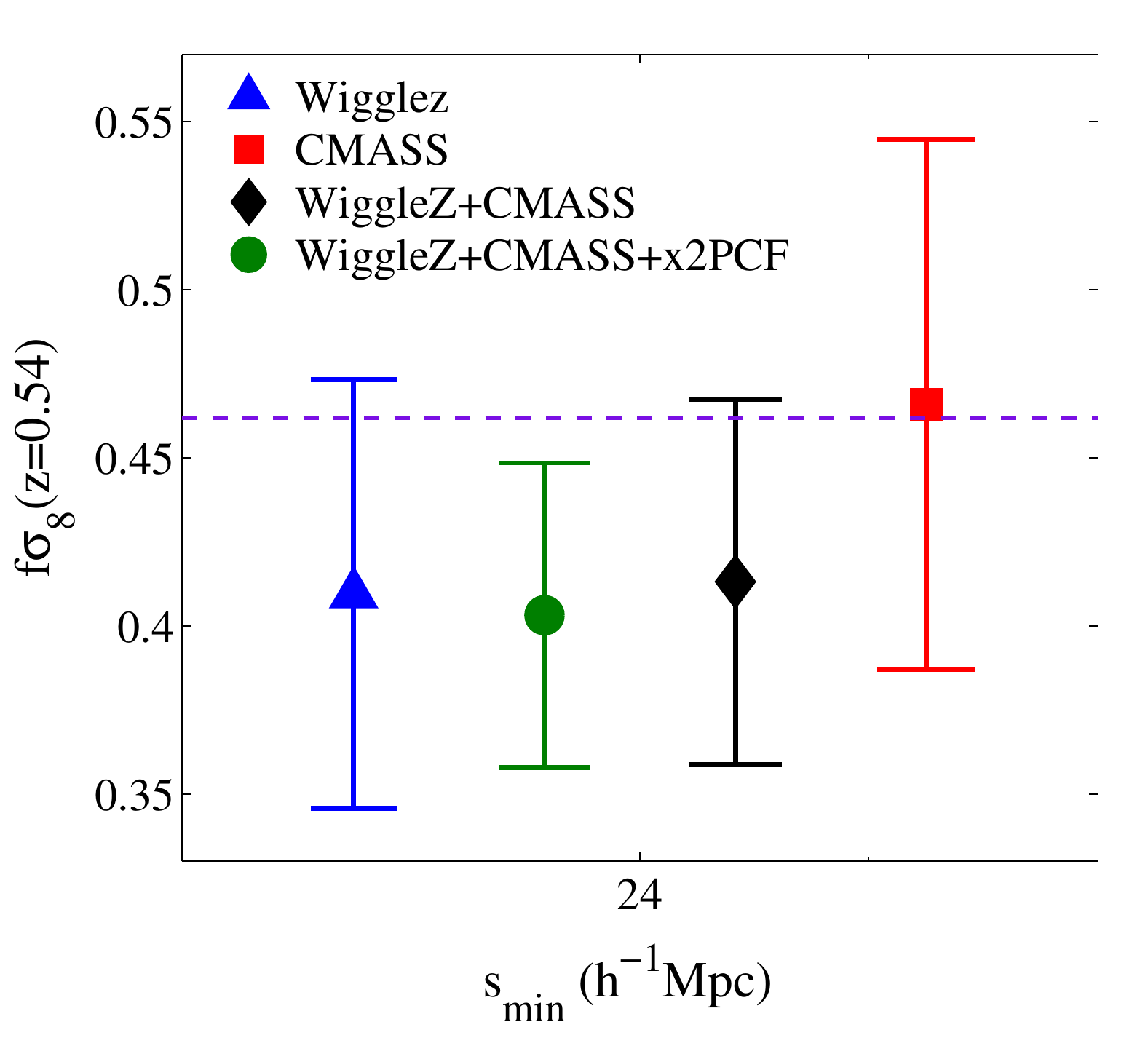}
\caption{\label{fig:f_fits}
Fits to the RSD model parameters using correlation multipoles with $24<s<80$ $h^{-1}$Mpc. We show 
results when analysing individual surveys and joint constraints. The purple line shows the prediction from 
WMAP5 cosmology.}
\end{figure}

Figure \ref{fig:f_fits} presents the parameter fits of $f\sigma_8(z=0.54)$ fitting the monopole and quadrupole of
the WiggleZ and CMASS auto- and cross-correlation on scales between
$s_{min}=24$ $h^{-1}$Mpc, and $s_{max}=80$ $h^{-1}$Mpc.
As shown in the previous section,  $s_{\rm min}=24$ $h^{-1}$Mpc is the minimum
scale where there are not important systematic deviations in the parameters from the study with the COLA mocks, and 
our fits to the observed data follow this trend. Table \ref{tab:results} lists the results for the parameter fits. 

Comparing the single-tracer fits for WiggleZ-BW and CMASS-BW galaxies, 
 there is agreement at the $1\sigma$ level for the values of $f\sigma_8(z=0.54)$, meaning that
when fitting to these scales there is evidence of no systematics depending on the type of galaxy used. 
Our constraints on the growth rate are consistent with our fiducial cosmology
$f\sigma_8(z=0.54)=0.46$.

Consistent with previous work,
we recover that the bias of the WiggleZ-BW galaxies, $b_W\sim 1$, is smaller than that of the
 CMASS-BW galaxies, $b_C\sim2$.
 The value of the best-fitting chi-squared statistic indicates that the model provides a reasonable fit to the data in all cases. 
 For the pairwise dispersion, values for the different tracers are consistent 
 with the predicted value from theory (section 4.1).

Combining the two tracers including their cross-covariance
yields slightly better constraints for $f\sigma_8(z=0.54)$ at the 
level of 10\% (compared to WiggleZ constraints alone). 
This result indicates that for these tracers, in a low density regime (where the common cosmic variance cancellation
does not improve the constraints, see Blake et. al 2013), 
even in the presence of a slightly larger Hartlap-Percival correction,
the improvement is due to reduced shot noise.  
When including the cross-correlations the improvement is of the order of 20\% 
(again, compared with WiggleZ constraints alone). 
In the case when we include the cross-correlations, we obtain our poorest value for $\chi^2/$d.o.f., 
implying that our simple constant $r_\times$ model may not describe all of the complexities of the cross-correlation.
Given this result,  we quote as result of our paper for the growth rate constraint the one obtained when 
we combine only the auto-correlations, yielding $f\sigma_8(z=0.54) = 0.413\pm0.054$.

\begin{table*}
\centering
\caption{\label{tab:results} Fits to the RSD model parameters when using multipoles with $24<s<80$ $h^{-1}$Mpc assuming WMAP5 $\sigma_8(z=0)=0.812$}
 \begin{tabular}{l c c c c c c c}
 \hline
 Tracers & $b_W\sigma_8(z=0.54)$ & $b_C\sigma_8(z=0.54)$  & $f\sigma_8(z=0.54)$ & $\sigma_v$ (kms$^{-1}$) & $r_\times$& $\chi^2$/d.o.f & d.o.f. \\
 \hline
  WiggleZ only                      & 0.651$\pm$0.046 &   -          & 0.409$\pm$0.059 & 205$\pm$144 & - & 1.11 &  28 - 4  \\
  CMASS only                      &    -               & 1.204$\pm$0.062 & 0.466$\pm$0.074 & 130$\pm$116 & - & 1.43 & 28 - 4\\
  WiggleZ+CMASS		  & 0.646$\pm$0.043& 1.233 $\pm$0.054   & 0.413$\pm$0.054 & 117$\pm$113 & -  & 1.28& 56 - 5\\
 WiggleZ+CMASS+x2PCF & 0.648$\pm$0.038  & 1.242$\pm$0.043  & 0.403$\pm$0.048 & 88$\pm$104 & 0.93 $\pm$ 0.03& 1.57 & 84 - 6\\
\hline
\end{tabular}
\end{table*}

\section{Summary \& Conclusions}

\begin{figure}
\includegraphics[width=8.5cm]{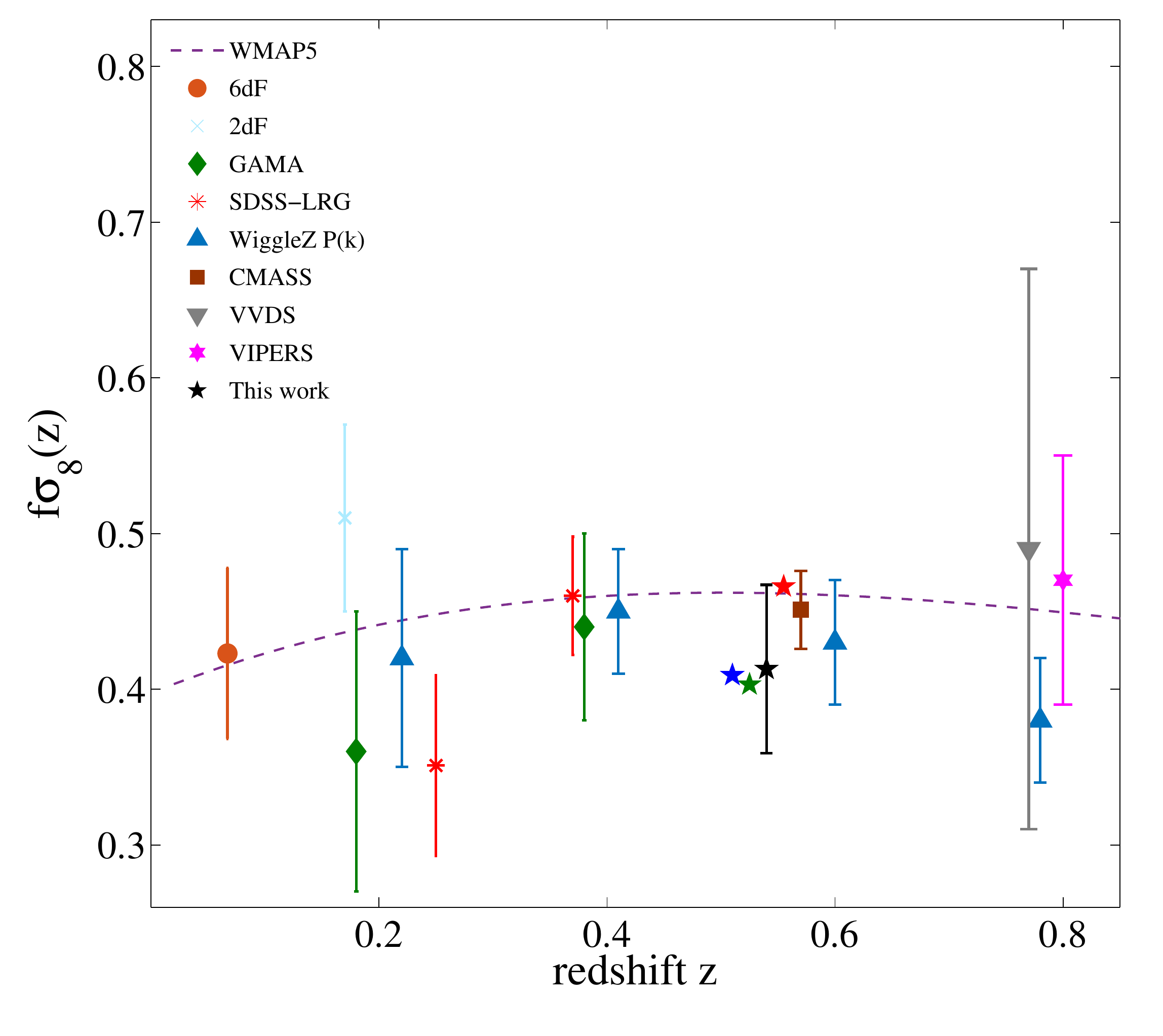}
\caption{\label{fig:compare_f_fits}
Fits to the growth rate $f\sigma_8(z)$ from different galaxy surveys: 6dF \citep{beutler_etal:12rsd},
2dFGRS \citep{hawkins_etal:03}, GAMA \citep{blake_etal:13}, WiggleZ \citep{blake_etal:11f}, 
SDSS LRGs \citep{samushia_etal:12}, CMASS-DR9 \citep{reid_etal:12}, VVDS\citep{guzzo_etal:08},
and VIPERS \citep{delatorre_etal:13}.}
\end{figure}

In this work we have presented the first cosmological RSD analysis using
data from two overlapping surveys, WiggleZ and CMASS.
After defining the overlap volumes, we measured 2-point auto- and cross-correlations
functions of these tracers;   after obtaining their multipoles and calculating their cross-covariance
using N-body mock catalogs, 
we compared them  with  RSD models in order to measure the growth rate of structure $f\sigma_8$ 
at an effective redshift
$z=0.54$. Our main findings are:

\begin{itemize}

\item The cross-correlation coefficient $r_\times$ between the WiggleZ-BW and CMASS-BW galaxies agrees 
with the expectation that
on large scales, the two classes trace similarly the large
scale structure with $r_\times\sim 1$. On smaller scales $s\lesssim20$ $h^{-1}$Mpc,
 $r_\times < 1$, likely produced by a combination
of a number of factors such as non-linear pairwise velocities, non-linear bias and stochasticity.

\item We tested redshift-space distortion models in mock catalogues simulating  WiggleZ and CMASS galaxies, 
including the selection functions of our  overlapping volumes.
When fitting scales $s > 24$ $h^{-1}$Mpc we recover our fiducial cosmological parameters using
different tracers, and that a single velocity dispersion provides an adequate description 
for the distortions in our range of  scales. We confirmed a lack of 
a significant improvement when using the multitracer technique, given the sparsity of the sampling
for these tracers. 

\item  The fits to $f\sigma_8(z)$ from all tracers are consistent with
each other and  with the predictions of a $\Lambda$CDM universe,
 showing no evidence for strong modelling systematic errors as a function of galaxy type. 
\end{itemize}

As shown in Figure \ref{fig:compare_f_fits},
 our combined fit for the growth rate $f\sigma_8(z=0.54)=0.413\pm0.054$
 is in excellent agreement with estimates from different surveys.  
Although more sophisticated models for the RSD can be employed, the motivation for our work
was to show consistency in the cosmological fits when using different tracers. 
This agreement provides further strong evidence for the robustness in the growth rate measurements which are
important for answering  the outstanding questions on the nature of dark energy
and large-scale gravity.

\section*{}

We thank  Ariel S\'anchez, H\'ector Gil-Mar\'in, 
Tamara Davis, David Parkinson, Ra\'ul Angulo, Andrew Johnson, Luis Torres, 
Shahab Joudaki and Caitlin Adams,
 for enlightening discussions and comments to 
this work. 
FM, CB, EK, JK were
supported by the Australian Research Council
Centre of Excellence for All-Sky Astrophysics (CAASTRO)
through project number CE110001020. CB acknowledges
the support of the Australian Research Council
through the award of a Future Fellowship. This work was
performed on the gSTAR national facility at Swinburne University
of Technology. gSTAR is funded by Swinburne and
the Australian Governments Education Investment Fund.
This research has made use of NASA's Astrophysics Data System.

Funding for SDSS-III has been provided by the Alfred P. Sloan Foundation, the Participating Institutions, 
the National Science Foundation, and the U.S. Department of Energy. 
SDSS-III is managed by the Astrophysical Research Consortium for the Participating Institutions of the 
SDSSIII Collaboration including the University of Arizona, the Brazilian Participation Group, Brookhaven National 
Laboratory, University of Cambridge, Carnegie Mellon University, University of Florida, 
the French Participation Group, the German Participation Group, Harvard University, 
the Instituto de Astrof\'isica de Canarias, the Michigan State/Notre Dame/JINA Participation Group, 
Johns Hopkins University, Lawrence Berkeley National Laboratory, Max Planck Institute for Astrophysics, 
Max Planck Institute for Extraterrestrial Physics, New Mexico State University, New York University, 
Ohio State University, Pennsylvania State University, University of Portsmouth, Princeton University,
 the Spanish Participation Group, University of Tokyo, University of Utah, Vanderbilt University, University of Virginia, 
 University of Washington, and Yale University.

\bibliography{wzboss2_rsd_fmarin}

\end{document}